\documentclass[epsf]{aa}
\usepackage{natbib}
\usepackage{epsfig}

\def\kms{\,km\,s$^{-1}$}       
\def\vsini{$v$\,sin\,$i$}      
\def\teff{$T_{\rm eff}$}
\def\logg{$\log g$}
\def\vsini{$v \sin i$}
\def\mictrb{$\xi_{\rm t}$}
\def\ftot{$f_{\oplus}$}
\def\kms{km\, s$^{-1}$}

\begin{document}
\title{Improved parameters for the transiting hot Jupiters WASP-4b and WASP-5b \thanks{Based on data collected with the FORS2 imager at the VLT-UT4 telescope and with the UVES spectrograph at the VLT-UT2 telescope (Paranal Observatory, ESO, Chile) in the programme 280.C-5003.}}
\subtitle{}
\author{M. Gillon$^{1}$, B. Smalley$^2$, L. Hebb$^3$, D. R. Anderson$^2$, A. H. M. J.  Triaud$^1$, C. Hellier$^2$, P. F. L.  Maxted$^2$, D. Queloz$^1$, D. M. Wilson$^2$}    

\offprints{michael.gillon@obs.unige.ch}
\institute{$^1$  Observatoire de Gen\`eve, Universit\'e de Gen\`eve, 51 Chemin des Maillettes, 1290 Sauverny, Switzerland\\
$^2$ Astrophysics Group, Keele University, Staffordshire, ST5 5BG, UK\\
$^3$ School of Physics and Astronomy, University of St. Andrews, North Haugh, Fife, KY16 9SS, UK\\}

\date{Received date / accepted date}
\authorrunning{M. Gillon et al.}
\titlerunning{Improved parameters for WASP-4b and WASP-5b}
\abstract{The gaseous giant planets WASP-4b and WASP-5b are transiting 12-magnitude solar-type stars in the Southern hemisphere. The aim of the present work is to refine the parameters of these systems using high cadence VLT/FORS2 $z$-band transit photometry and high-resolution VLT/UVES spectroscopy. For WASP-4, the  new estimates for the planet radius and mass from a combined analysis of our VLT data with previously published transit photometry and radial velocities are $R_p = 1.30^{+0.05}_{-0.04}$ $R_J$ and $M_p = 1.21^{+0.13}_{-0.08}$ $M_J$, resulting in a density  $\rho_p = 0.55^{+0.04}_{-0.02}$ $\rho_J$. The radius and mass for the host star are $R_\ast = 0.87^{+0.04}_{-0.03}$ $R_\odot$ and $M_\ast = 0.85^{+0.11}_{-0.07}$  $M_\odot$.  Our ground-based photometry reaches 550 ppm at time sampling of $\sim$ 50 seconds.  Nevertheless, we also report  the presence of an instrumental effect on the VLT that degraded our photometry for the WASP-5 observations. This effect could be a major problem for similar programs.  Our new estimates for the parameters of the WASP-5 system are $R_p = 1.09 \pm 0.07$ $R_J$, $M_p = 1.58^{+0.13}_{-0.10}$ $M_J$, $\rho_p = 1.23 ^{+0.26}_{-0.16}$ $\rho_J$, $R_\ast = 1.03^{+0.06}_{-0.07}$ $R_\odot$,  and $M_\ast = 0.96^{+0.13}_{-0.09}$ $M_\odot$. The measured size of WASP-5b agrees well with the basic models of irradiated planets, while WASP-4b is clearly an `anomalously' large planet. 
 }

\maketitle

\section{Introduction}

So far, the planets that transit their parent stars have  undoubtedly brought the most important pieces of information about the physics and composition of the planetary objects outside our Solar System (see review by Charbonneau et al. 2007). Most of the transit detections are due to a few ground-based wide-field surveys targeting stars brighter than V$\sim$13: HAT (Bakos et al. 2004), SuperWASP (Pollaco et al. 2006), TrES (O'Donovan et al. 2006), and XO (McCullough et al. 2005). Among these surveys,  SuperWASP is the one showing the largest harvest so far. This efficiency is not only due to the constant optimization of the observational and follow-up strategy, reduction and data analysis (Cameron et al. 2007), but also to the recent starting of the Southern counterpart of the SuperWASP-North facility. Located at the Sutherland Station of the South African Astronomical Observatory, SuperWASP-South brings a second field of view of 482 square degrees to the survey, allowing it to search for transiting planets in a large portion of the sky. 

The planets WASP-4b (Wilson et al. 2008, hereafter W08) and WASP-5b (Anderson et al. 2008, hereafter A08) were the first transiting planets detected by SuperWASP-South. They are both gas giants slightly heavier than Jupiter and orbiting very close (P  = 1.338 and 1.628 days) to 12 mag solar-type stars. The analysis of the WASP and follow-up data led to radius values of about 1.4 and 1.1 $R_J$. With an estimated density $\sim 0.4$ $\rho_J$, WASP-4b appeared to belong to the subgroup of the planets with a radius larger than predicted by basic models of irradiated planets (Burrows et al. 2007a, Fortney et al. 2007) unlike WASP-5b ($\rho \sim 1.2$ $\rho_J$).  WASP-4b is slightly less massive than WASP-5b (1.2 $vs$ 1.6 $M_J$), while both planets have a similar irradiation (1.89 $vs$ 1.92 10$^9$ erg s$^{-1}$ cm$^{-2}$), semi-major axis (0.023 $vs$ 0.027 AU) and host star spectral type (G7V $vs$ G4V).

Several hypotheses have been proposed to explain the radius anomaly shown by some highly irradiated planets such as WASP-4\,b (see Guillot 2008), most importantly tides (Bodenheimer et al. 2001; Jackson et al. 2008),  tides with atmospheric circulation (Guillot \& Showman 2002) and enhanced opacities (Guillot et al. 2006, Burrows et al 2007a). Receiving similar irradiation from their host stars while having significantly different radii, WASP-4b and WASP-5b represent a good test for theory and an interesting opportunity of progress on our understanding of the radius heterogeneity observed among the highly irradiated planets. It is thus desirable to obtain for these two planets the highest precision possible on the system parameters. This motivated us to use the VLT to obtain (1) a high cadence high precision transit light curve with the FORS2 camera, and (2) a high resolution spectrum of the host stars with the UVES spectrograph.  We present respectively in Sec. 2 and 3 these new VLT photometric and spectroscopic observations and their reduction. We analyze these new data in combination with former transit photometry and RV measurements in Sec. 4. The results of our analysis are discussed in Sec. 5. 

\section{VLT/FORS2 Transit Photometry}

\subsection{WASP-4}

The photometry for WASP-4 was obtained on October 23, 2007. Altogether 339 exposures were acquired with the FORS2 camera on the VLT/UT4 telescope from 01h01 to 06h50 UT. To have enough reference flux to properly correct the photometry from atmospheric effects, the standard resolution mode was used, resulting in a 6.8' $\times$ 6.8' field of view. To obtain a good time sampling of the light curve, a 2$\times$2 binning of the pixels was performed, resulting in a pixel scale of  0.25''/pixel. The exposure time was tuned by the ESO staff astronomer to 20s while the mean read-out plus overhead time was 34s. We chose to observe in the $z$-GUNN+78 filter ($\lambda_{eff}= 910 $ nm, FWHM = 130.5 nm)  to minimize the impact of the stellar limb-darkening uncertainty on the deduced system parameters. A very large defocus was used to obtain a good duty cycle and to minimize the influence of flat-fielding errors: the mean characteristic profile width was 50 pixels = 12.5''.  Despite this large defocus, there was no PSF overlap for the target and the reference stars. The guiding system was turned on to make the stellar fluxes registered on nearly the same pixels during the run (centroid jitter $\sim$ 3.5 pixels for the whole run). There was an interruption of  25 minutes in the sequence due to a technical problem with the secondary mirror setting, fortunately before the transit. The quality of the night was photometric. The moon illumination was 87 \%. It was at $44^{\circ}$ at closest from the target. The airmass decreased from 1.08 to 1.05 then increased to 1.95 during the run (Fig. 1). The defocus was tuned several times to adapt it to atmospheric transparency variations due to the increase of airmass. 

After a standard pre-reduction, the stellar fluxes were extracted for all the images with the {\tt IRAF DAOPHOT} aperture photometry software (Stetson, 1987). As the defocus was not the same for the whole run, the reduction parameters were adapted to the characteristic profile width of each image. 

The transit is already very clear in the absolute flux curve of WASP-4. But as shown in Fig. 2, the absolute  photometry of WASP-4 and of several other stars in the field suffer from an unexpected effect: while the shape of the largest part of these curves shows a nice airmass-flux correlation indicating that the night transparency conditions were very good, the first part seems to be affected by a large systematic dependent
 on the position on the chip. High-accuracy transit photometry has already been obtained with the FORS cameras (e.g. Gillon et al. 2007a, Pont et al. 2007), and this systematic was not detected in these former data. The main difference  between these former observations and ours is the large defocus that we used. The explanation that we and the ESO staff favor  is  linked to the strange shape taken by the primary mirror M1 with respect to the secondary M2 in case of out-of-focus observations, with variations induced by the different amount of tangential component of the gravity as the dominant effect. Indeed, the active optics system of the telescope is supposed to compute for each exposure an optimal shape for the M1 so to correct for tangential gravity pull, and actuators perform micrometrical shifts of the M1 to obtain the computed shape, but here the active optics system was turned off at the beginning of the run to obtain the required huge defocus. Thus the needed correction for the tangential gravity pull was not applied. The resulting spatial difference in illumination of the chip could then have been rather large at the beginning of the run, when the telescope was close to the meridian. As it moved away from meridian, the tangential correction became less  important, and so did the effect.  

Differential photometry was performed using the flux of several bright stable stars in the field, but the obtained curve is still plagued with a large systematic in its first part (see Fig. 3). Fortunately, the transit occured in the second part of the run for which the effect seems to  be absent. We thus decided to use only the data after BJD = 2454396.625, for which the photometry seems to be reliable and accurate. 

After a careful selection of the reference stars and the reduction parameters, we subtracted a linear fit for magnitude $vs$ airmass  to correct the photometry for differential reddening using the out-of-transit (OOT) data. The corresponding fluxes were then normalized using the OOT part of the photometry. The resulting transit light curve is shown in Fig. 4, with the best-fit transit model (see Sec. 4) superimposed. The $rms$ of the first OOT part is 420 ppm. This  value is very close to the theoretical error per point obtained from the photon noise of the target and the reference stars, the sky background, read-out and scintillation noises (Gilliland et al. 1993): 400 ppm. For the second OOT part, the measured $rms$ is 740 ppm while the median theoretical error is 510 ppm. This largest discrepancy between both values comes probably from the amplification of the effect of any transparency inhomogeneity across the field at high airmass. Indeed, the airmass ranges from 1.45 to 1.95 in the second OOT part.

The $rms$ and the time sampling are not the only parameters needed to evaluate the quality of a photometric time series, the level of low-frequency noise (red noise) has also to be taken into account, especially for high SNRs (Pont et al. 2006). We estimated the level of red noise $\sigma_r$ in our photometry using the equation (Gillon et al. 2006):
\begin{equation}\label{eq:g}
\sigma_r =  \bigg(\frac{N\sigma_N^2 - \sigma^2}{N - 1}\bigg)^{1/2}\textrm{,}
\end{equation}
\noindent
where $\sigma$ is the $rms$ in the original OOT data and $\sigma_N$ is the standard deviation after binning the OOT data into groups of $N$ points. We used $N = 20$, corresponding to a bin duration similar to the ingress/egress timescale. The obtained value for $\sigma_r$ is quite small: 110 ppm. 

\begin{figure}
\label{fig:b}
\centering                     
\includegraphics[width=8cm]{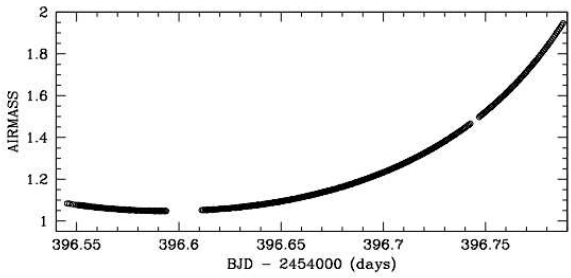}
\caption{Evolution of the airmass during the WASP-4 VLT/FORS2 run.}
\end{figure}

\begin{figure}
\label{fig:a}
\centering                     
\includegraphics[width=9cm]{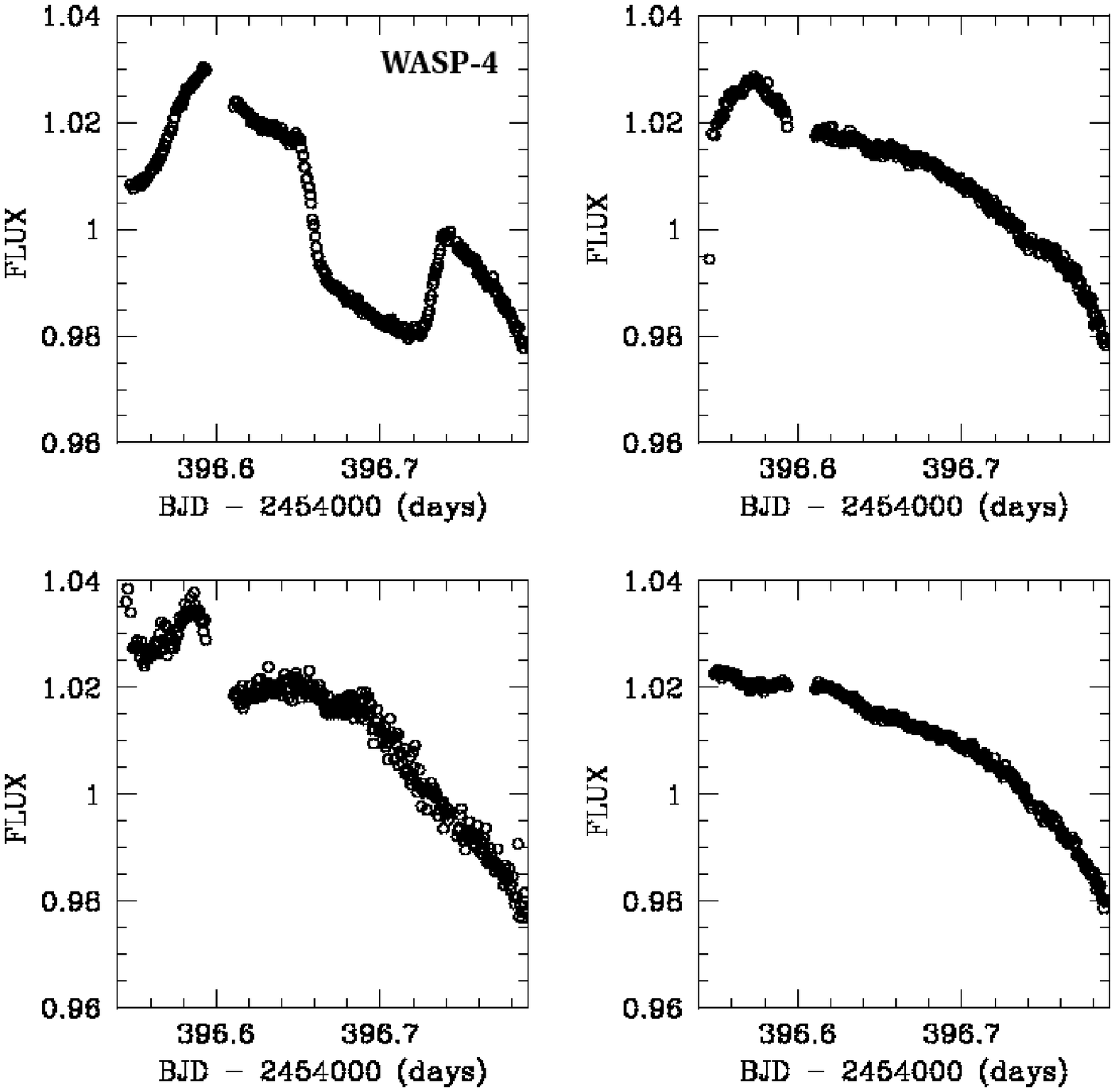}
\caption{$Top$ $left$ : normalized absolute VLT/FORS2 photometry for WASP-4. $Top$ $right$ $and$ $bottom$: normalized absolute photometry for other stars of the field.  }
\end{figure}

\begin{figure}
\label{fig:c}
\centering                     
\includegraphics[width=9cm]{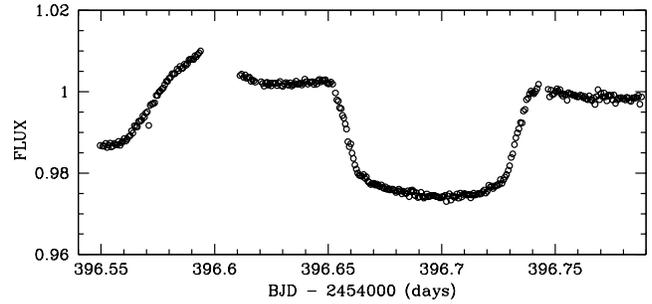}
\caption{Differential photometry for WASP-4 before rejection of the part of the curve damaged by the systematic and differential reddening correction.}
\end{figure}

\begin{figure}
\label{fig:d}
\centering                     
\includegraphics[width=9cm]{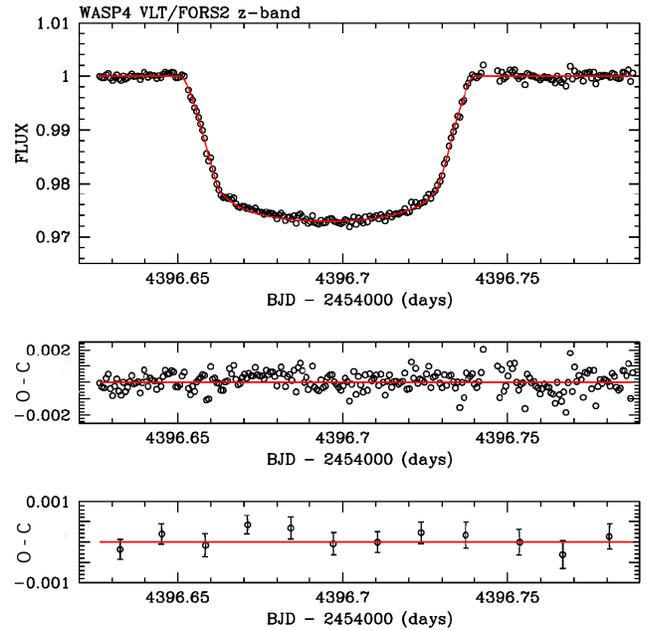}
\caption{$Top$: VLT/FORS2 $z$-band transit photometry for WASP-4. The best-fit transit curve is superimposed in red. $Middle$: residuals of the fit ($rms$ = 570 ppm ). $Bottom$: residuals of the fit after binning per 20 points ($rms$ = 200 ppm).}
\end{figure}

\subsection{WASP-5}

The photometry for WASP-5 was obtained on November 15, 2007. 337 exposures were acquired from 00h24 to 05h45 UT, again with the FORS2 camera. The same observational strategy as for WASP-4 was used. The quality of the night was photometric. The moon was in its first quarter (illumination = 24 \%), and its closest distance to the target was  $48^{\circ}$  at the end of the run. The airmass decreases from 1.06 to 1.04 then increased to 1.97 (Fig. 5). 

The same reduction procedure as for WASP-4 was used. Here again, the photometry of the first part of the run shows the presence of a systematic at the percent level (Fig. 6). We notice that the amplitude of the effect depends here too on the position on the chip and appears to affect only the low-airmass photometry, it is thus probably the same effect than for WASP-4 (see above). Unfortunately, the transit happened at the beginning of the run and its photometry is too damaged to be useful. We thus decided to use only the UVES spectroscopy (see Sec. 3) and former data to characterize this system.

\begin{figure}
\label{fig:f}
\centering                     
\includegraphics[width=8cm]{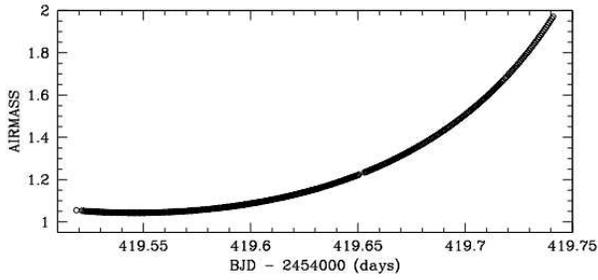}
\caption{Evolution of the airmass during the WASP-5 VLT/FORS2 run.}
\end{figure}

\begin{figure}
\label{fig:e}
\centering                     
\includegraphics[width=9cm]{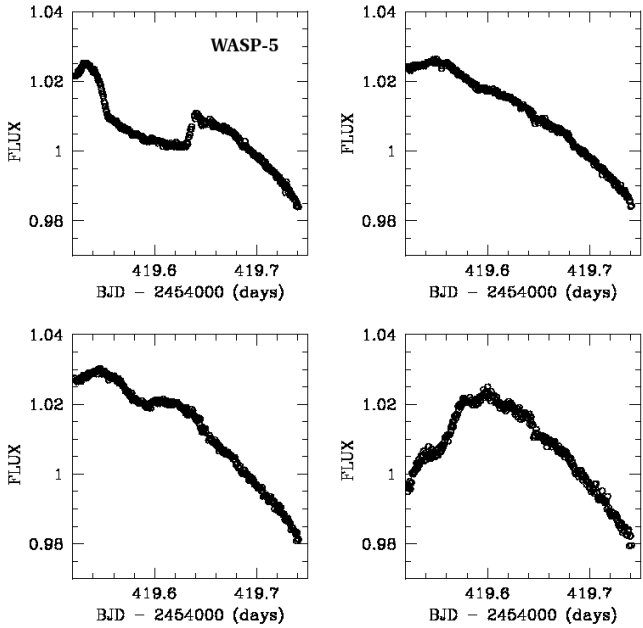}
\caption{$Top$ $left$ : normalized absolute VLT/FORS2 photometry for WASP-5. $Top$ $right$ $and$ $bottom$: normalized absolute photometry for other stars in the field.  }
\end{figure}

\section{VLT/UVES Spectroscopy}

High-resolution spectra of WASP-4 and WASP-5 were obtained using the UVES spectrograph on the VLT/UT2 telescope on November 27, 2007. The red arm was used with standard setting 580, giving spectral coverage from 4780 to 6808 \AA, except from 5758 to 5833 \AA\ due to the gap between to two CCD detectors. A 1-arcsec slit was used giving a spectral resolution of 40\,000. Exposure times of 600\,s were used yielding mean S/N of 94:1 and 102:1 for WASP-4 and WASP-5, respectively. The standard pipeline reduction products were used in the analysis.

The analysis was performed using the {\sc uclsyn} spectral synthesis package (Smith 1992; Smalley et al. 2001) and {\sc atlas9} models without convective overshooting (Castelli et al. 1997). The $H_\alpha$ line was used to determine the effective temperature (\teff), while the Na {\sc i} D and Mg {\sc i} lines were used as surface gravity (\logg) diagnostics. The parameters obtained from the analysis are listed in Table~1.

The equivalent widths of several clean and unblended lines were measured. Atomic line data was mainly taken from the Kurucz \& Bell (1995) compilation, but with updated van der Waals broadening coefficients for lines in Barklem et al. (2000) and $\log gf$ values from Gonzalez \& Laws (2000), Gonzalez et al. (2001) or Santos et al. (2004). A value for microturbulence (\mictrb) was determined from Fe~{\sc i} using Magain's (1984) method. The ionization balance between Fe~{\sc i} and Fe~{\sc ii} and the null-dependence of abundance on excitation potential were used as additional  \teff\ and \logg\   diagnostics (Smalley 2005).

In addition to the spectral analysis, we have also used published broad-band photometry to estimate the total observed bolometric flux (\ftot). For WASP-4, TYCHO-2 (H{\o}g et al. 2000), USNO-B1.0 R-mag. (Monet et al. 2003) and 2MASS (Cutri et al. 2003) were used, while for WASP-5 GALEX nuv flux (Morrissey et al. 2007), NOMAD (Zacharias et al. 2004), DENIS (Fouqu\'{e} et al. 2000) and 2MASS were used, but no TYCHO photometry was available. The photometry was converted to fluxes and the best-fitting Kurucz (1993) model flux distribution found, which was integrated to determine \ftot. The Infrared Flux Method (Blackwell \& Shallis 1977) was then used with 2MASS magnitudes to determine \teff\ and stellar angular diameter ($\theta$). The results are given in Table~1 and are consistent with those from the spectroscopic analysis. 

We have determined the elemental abundances of several elements (listed in Table~1) from their measured equivalent widths. The quoted error estimates include those given by the uncertainties in \teff, \logg\ and \mictrb, as well as the scatter due to measurement and atomic data uncertainties. In our spectra the Li {\sc i} 6708\AA\ line is not detected. Thus we can only give upper-limits on the Lithium abundances. 

\begin{center}
\begin{table*}[h]
\begin{tabular}{c|cc}

             & WASP-4             & WASP-5             \\
\hline
\teff        & 5500 $\pm$ 100 K   & 5700 $\pm$ 100 K   \\
\logg        & 4.5 $\pm$ 0.2      & 4.5 $\pm$ 0.2      \\
\mictrb      & 1.1 $\pm$ 0.2 \kms & 1.2 $\pm$ 0.2 \kms \\
\vsini      & 2.0 $\pm$ 1.0 \kms       & 3.5 $\pm$ 1.0  \kms        \\
{[Fe/H]}     & $-0.03$ $\pm$ 0.09 & +0.09 $\pm$ 0.09    \\
{[Si/H]}     &   +0.08 $\pm$ 0.05 & +0.21 $\pm$ 0.06    \\
{[Ca/H]}     &   +0.04 $\pm$ 0.14 & +0.07 $\pm$ 0.14    \\
{[Sc/H]}     &   +0.07 $\pm$ 0.13 & +0.21 $\pm$ 0.10    \\
{[Ti/H]}     &   +0.11 $\pm$ 0.12 & +0.14 $\pm$ 0.08    \\
{[V/H]}      &   +0.09 $\pm$ 0.07 & +0.16 $\pm$ 0.07    \\
{[Cr/H]}     &   +0.06 $\pm$ 0.08 & +0.07 $\pm$ 0.10    \\
{[Co/H]}     &   +0.07 $\pm$ 0.07 & +0.20 $\pm$ 0.07    \\
{[Ni/H]}     &   +0.02 $\pm$ 0.10 & +0.15 $\pm$ 0.05    \\
log N(Li)    & $<$ 0.8            & $<$ 0.5            \\
\teff(IRFM)  & 5470 $\pm$ 130 K &  5740 $\pm$ 130 K \\
$\theta$(IRFM)& 0.031 $\pm$ 0.002 mas&0.032 $\pm$ 0.002 mas \\
\hline
\end{tabular}
\caption{Spectroscopic parameters obtained in this work for WASP-4 and WASP-5.}
\label{wasp-params}
\end{table*}
\end{center}

\section{Data Analysis}

\subsection{Determination of the system parameters}

\subsubsection{WASP-4}

We derived stellar and planetary parameters for the system by fitting simultaneously our new VLT
$z$-band transit light curve with the data presented in W08, i.e.\ (1) the WASP R-band photometry,
(2) the FTS i-band transit light curve, (3) an EulerCAM R-band transit light curve and (4)
14 CORALIE radial velocity measurements.

The data were used as input into the Monte
Carlo Markov Chain (MCMC) code described in Cameron et al. (2007)
which was designed specifically to solve the multi-variate problem of transiting star-planet systems.   
Via the MCMC approach, the fitting code repeatedly adopts trial parameters until
it converges on a set of values which produces the best model velocity
curve and model light curves.  In short, nine parameters were used to
describe the light curves and radial velocity curve of the host star
including the orbital period P, the time of minimum light T$_0$,
the transit depth $\delta$, the total transit duration $t_T$, the impact parameter $b$,
the stellar mass $M_{*}$, the stellar velocity amplitude $K_1$, the
systemic radial velocity $\gamma$, the orbital eccentricity $e$,
and the longitude of periastron $\omega$.  These nine fitted parameters 
determine the physical properties of the star-planet system, including the
masses and radii of the star and planet and the orbital inclination and separation.  
The goodness-of-fit statistic used to assess 
the best parameters is the sum of the $\chi^2$ for all the data curves 
with respect to the models.  Model light curves were derived according to the formalism outlined in
Mandel \& Agol (2002), adopting the small-planet approximation and using the non-linear limb 
darkening coefficients from Claret (2000, 2004) for the appropriate
photometric filters.  The code also has the option to apply a Bayesian main sequence 
prior on the stellar mass and radius which acts to keep the 
star on the main sequence.  However, due to the exceptional quality
of the follow-up photometry, we did not apply this constraint in
the analysis of WASP-4.  

We ran the MCMC code in an iterative fashion to derive the 
best overall solution for the properties of the star-planet system.
We also combined the physical properties derived from the MCMC code 
with the stellar parameters from the spectral synthesis to determine 
the evolutionary status of the host star and to confirm that 
the stellar properties, including mass (which is not directly measured), were consistent 
with each other and with theoretical stellar evolution models.

In the initial run of the MCMC code, we adopted initial guesses for the light curve 
parameters from the results of the box-least squares analysis
of the WASP data.  We also assumed a starting value for the eccentricity
of 0.02, a systemic RV equal to the mean of the velocity data, and a velocity
amplitude derived by fitting a sinusoidal velocity variation to the observed
RVs by minimizing $\chi^2$.  The initial guess for the stellar mass
was derived by interpolating the zero age main sequence, solar metallicity stellar evolution 
isochrone of Girardi et al. (2000) at the temperature of the host star.
In addition, the input light curves had theoretically 
derived uncertainty values for each photometric measurement.   The uncertainties were 
computed by considering shot noise, scintillation, read-out and background noises, but did not 
include correlated noise.  For the RV measurements, the MCMC code adds quadratically an additional 
RV jitter to the theoretical uncertainties so that the reduced $\chi^2$ of the RV data compared to 
the model curve is approximately equal to 1.  Finally, we allowed the eccentricity to float freely.  

The results of this initial run were 
used to inform the second and final run of the code in the following ways.
First, the initial eccentricity result was well within 1$\sigma$
of zero.  Thus, for final run, we fixed the eccentricity
to zero, assuming a circular orbit for the planet.
Next, the best fitting model light curves from the initial run were used
to make a correlated noise measurement for each photometric
time-series.  We employed the method described in Sec. 2.1 (Eq. 1),
but using the entire residual time-series, rather than just the OOT parts.
Table~2 presents the red noise values obtained.   
For each photometric measurement, we added the appropriate correlated noise 
value in quadrature to the theoretical uncertainty.  The resulting light
curves were input into the final run of the MCMC code.

Finally, the first MCMC run produced a measurement of the mean stellar density 
($\rho_s = M_s/R_s^3$) which we used to refine the initial
guess of the stellar mass.  We converted the derived stellar
density to $R_s/M_s^{1/3}$ in solar units, and compared this
property and the stellar temperature in a modified Hertzsprung-Russel (HR) to the
Girardi solar metallicity models.  The quantity, $R_s/M_s^{1/3}$,
depends only on the observed transit properties (duration, depth, impact parameter, and orbital period)
and is independent of the measured temperature.
We generated the same property from the mass and log $g$ values in the models,
and then interpolated the models in the $R/M^{1/3}$-Teff plane to determine
a mass and age for WASP-4.  We interpolated linearly along two consecutive
mass tracks to generate an equal number of age points between the
zero-age main sequence and the evolutionary state defined as Te-M which is the
stage where the star reaches core Hydrogen exhaustion.  We then interpolated
between the mass tracks along equivalent evolutionary points to find the
mass and age from the models that best match the stellar properties derived
from the MCMC code (density) and the spectral synthesis (temperature).  In this way,
we obtained a new intital guess for the stellar mass of, $M_{*}=0.93~M_{\odot}$.

We then implemented a second and final run of the MCMC code, (1) fixing $e=0$, 
(2) including correlated noise in the photometric uncertainties, 
(3) applying the results from the first MCMC run as the initial guesses
for the fitted parameters, and (4) adopting a prior on the stellar mass
of $M_{*}=0.93~M_{\odot}$.  The results of this run are given in Table~3.
The obtained value for the RV jitter is 7 m s$^{-1}$, identical to the value 
deduced in W08.

Lastly, we plotted the final stellar parameters on the modified HR
diagram and interpolated the model tracks as described above 
to determine the age of the system.  We derive an age of $5.2^{+3.8}_{-3.2}$~Gyr.
We also confirmed that the final mass derived in the MCMC code was consistent with
the observed temperature and the stellar evolution models (Figure 7).   
The MCMC result gives a mass of $0.85^{+0.11}_{-0.07}$ $M_{\odot}$ which is
within the 1$\sigma$ uncertainty on the mass determined from the
theoretical tracks ($0.93 \pm 0.05 M_{\odot}$).

We also made an independent analysis aiming to test if the quality of our VLT $z$-band
transit photometry is good enough to constraint reliably the limb-darkening of the star. In this 
analysis, we used as data our VLT  transit light curve with only the WASP photometry and the 
CORALIE radial velocities. We assumed a quadratic  limb-darkening law for the photometric
models to minimize the number of free parameters and allowed  the two limb-darkening
coefficients $u_1$ and $u_2$ to float for the VLT photometry only, using as jump parameters
not these coefficients themselves but the combinations $2 \times u_1 + u_2$ and $u_1 - 2 
\times u_2$ to ensure that the  obtained uncertainties are uncorrelated (Holman et al. 2006). 
We obtain values for the transit depth and duration that are slightly different from the values 
shown in Table 3, but all the deduced physical parameters are in good agreement with the 
previous values. The obtained values for the limb-darkening coefficients are $u_1 = 
0.299^{+0.006}_{-0.026}$ and $u_2 = 0.248^{+0.016}_{-0.023}$. These values are physically 
plausible in the sense that they produce a monotonically decreasing intensity from the center 
of the star to the limb, but they  are consistent with the values interpolated from Claret's tables 
(2000, 2004): $u_1$ = 0.266 and $u_2$ = 0.302. This disagreement could be due to the fact
that our VLT photometry has a small but significant level of correlated noise able to modify
slightly the actual shape of the transit.  The amplitude of this red noise is in fact similar to 
the difference between the models fitted with and without free limb-darkening coefficients. 
We thus prefer to consider the values for the system parameters obtained with a fixed  
non-linear limb-darkening law as our final ones. 

\begin{figure}
\label{fig:e}
\centering                     
\includegraphics[width=9cm]{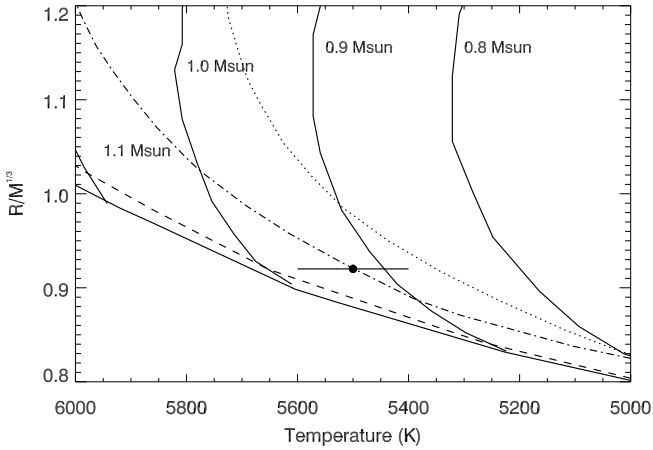}
\caption{Modified HR diagram showing 
$R/M^{1/3}$ in solar units versus effective temperature.
The properties of the of WASP-4 are overplotted 
on the theoretical stellar evolutionary models of Girardi et al. (2000).
The mass tracks are labeled and the isochrones are 100 Myr (solid), 
1 Gyr (dotted), 5 Gyr (dot-dashed), 10 Gyr (dotted).  According to
the models, the host star has an age of $5.2^{+3.8}_{-3.2}$ Gyr. 
The y-axis error bars are smaller than the data point.}
\end{figure}

\subsubsection{WASP-5}

An analysis similar to the one described above  was performed for WASP-5 using data presented in A08: (1) the WASP photometry, (2) a FTS $i$-band transit light curve, (3) an EulerCAM R-band transit light curve and (4) 11 CORALIE radial velocities.  As in A08,  no additional RV jitter was needed to obtain a final reduced $\chi^2$ close to 1 for the RV model. No Bayesian main sequence prior was used as for WASP-4. The orbital eccentricity obtained in the first  MCMC run was non-zero at the 2$\sigma$ level, so it was kept as free parameter in the second run, leading to a final value of  $0.04 \pm 0.02$ (see Table 3). Nevertheless, $e$ and $\omega$ are not orthogonal parameters because $e$ cannot be $<$ 0, and this asymmetry could lead to an overestimation of $e$ for low eccentricity orbit. To check that our marginal detection of a non-null eccentricity is reliable, we made a new analysis using as jump parameters  $e$.$\cos \omega$ and  $e$.$\sin \omega$, which are orthogonal,  and calculated $e$ and $\omega$ afterwards. We obtained $e = 0.049^{+0.020}_{-0.017}$ and $\omega =  0.73^{+0.30}_{-0.45}$, in very good agreement with the values presented in Table 3. We thus claim the marginal detection of a non-null eccentricity for WASP-5b.

WASP-5 is slightly super solar metallicity with [Fe/H]= $+0.09 \pm 0.09$ (see Table 1), so mass tracks of the Girardi models in metallicity at +0.09 were first linearly interpolated between the zero and +0.20 metallicity mass tracks before the obtained stellar parameters were plotted on the modified HR diagram and model tracks were interpolated (Fig. 8). From this final step, we derive an age of $5.4^{+4.4}_{-4.3}$ Gyr and a stellar mass of $1.00^{+0.07}_{-0.05} M_{\odot}$, within the error bar of the MCMC value of $0.96^{+0.13}_{-0.09}$ $M_{\odot}$ (see Table 3). 

\subsection{Transit timings}

Fixing all the system parameters except epoch to the ones deduced from the above analysis, we fitted with our MCMC code a transit profile to each  transit to obtain individual timings. For the WASP data, the transits of the season 2006 and 2007 were folded together to obtain a timing per season. Table 2 gives the obtained timings and error bars. From these results, the VLT $z$-band photometry for WASP-4 appears to show a formal error of $\sim$ 5s on its deduced timing, comparable to the best precision obtained from space (see e.g. Knutson et al. 2007). Such a precision is doubtful for our ground-based data, because we know that despite their high quality they  have a low but still significant level of covariant noise able to bring a systematic error on the deduced timing. To assess the influence of this covariant noise on our timing precision, we analyzed each transit with the method described in Gillon et al. (2007b), for which the estimation of the errors is based on the `prayer bead' procedure: after having determined the best-fitting eclipse model, a large number of fits are performed and for each of them the residuals of the initial fit are shifted sequentially about a random number and then added to the eclipse solution. This procedure allows to take into account the actual covariant noise of the data. At the end, the error bars of the fitted parameters are determined from the distribution of their derived values.  Table 2 presents also the values and error bars obtained with this method. We notice that for the data showing the largest level of covariant noise, the WASP data, the error bars on the timing is $\sim$4 times larger. For the Euler transits having a rather low level of covariant noise, the difference is much smaller. Interestingly, the error bars on the WASP-4 VLT transit timing is  3 - 4 times larger than the one obtained with the MCMC code and is only slightly better than the one of the Euler transits. This shows clearly that covariant noise has an important impact on the transit timing precision. While Table 2 presents the timings and errors obtained with both methods, we outline that our final results are the ones obtained with the `prayer bead' method: only these values should be used in future analysis. 

Figure 9 shows for both planets the residuals from the subtraction to these timings of the calculated transit timings based on ephemeris presented in Table 3.  The transits of WASP-4 show no clear sign of  period variability. A linear fit to the transit timings as a function of the transit epoch results in a  period of 1.3382319 $\pm$  0.0000068 days, in excellent agreement with the period obtained from our combined MCMC analysis. The reduced $\chi^2$ of the fit is 1.35, indicating that the epoch - timing relation is well modeled by a line. For WASP-5, the period deduced from a similar fit is 1.628430 $\pm$ 0.000013 days,  in good agreement with the value obtained from the combined analysis, but here the reduced $\chi^2$ is 5.7. At this stage, we cannot assign any firm significance to a possible period variability, because a shift of one of the two most precise timings (FTS or Euler) due to an unknown systematic effect could lead to a much bettered reduced $\chi^2$, but  obtaining more precise transit timings for WASP-5 is desirable.

\begin{table*}[h]
\begin{tabular}{cccc}
\hline
Light curve & $\sigma_r$ [ppm] & Mid-transit timing (MCMC) [BJD] &  Mid-transit timing ('prayer bead') [BJD] \\
\hline
WASP-4 WASP06 R-band & 3150 & $2453963.10863^{+0.00074}_{-0.00081}$ & $2453963.1086^{+0.0025}_{-0.0023}$ \\
WASP-4 WASP07 R-band & 4020 & $2454364.57722^{+0.00068}_{-0.00075}$ & $2454364.5757^{+0.0021}_{-0.0033}$ \\
WASP-4 Euler R-band & 0 & $2454368.59244^{+0.00022}_{-0.00019}$ & $2454368.59266^{+0.00025}_{-0.00027}$ \\
WASP-4 FTS i-band& 510 & $2454371.26812^{+0.00033}_{-0.00028}$ & $2454371.26738^{+0.00097}_{-0.00087}$ \\
WASP-4 VLT z-band & 190 & $2454396.695410 \pm 0.000051$ & $2454396.69548^{+0.00015}_{-0.00026}$ \\
\\
WASP-5 WASP06 R-band & 3260 & $2453945.71962^{+0.00091}_{-0.00093}$ & $2453945.7187^{+0.0041}_{-0.0028}$ \\
WASP-5 WASP07 R-band & 1490 &  $2454364.2283^{+0.0012}_{-0.0013}$ & $2454364.2285^{+0.0057}_{-0.0069}$ \\
WASP-5 Euler R-band & 360 & $2454383.76684^{+0.00025}_{-0.00024}$ & $2454383.76738^{+0.00031}_{-0.00032}$ \\
WASP-5 FTS i-band& 790 &  $2454387.02221^{+0.00034}_{-0.00037}$ & $2454387.02197^{+0.00071}_{-0.00050}$ \\
\hline
\end{tabular}
\caption{Typical red noise values and mid-transit times evaluated for the photometric time-series used in this analysis.}
\label{red noise}
\end{table*}

\begin{figure}
\label{fig:g}
\centering                 
\includegraphics[width=9cm,angle=0]{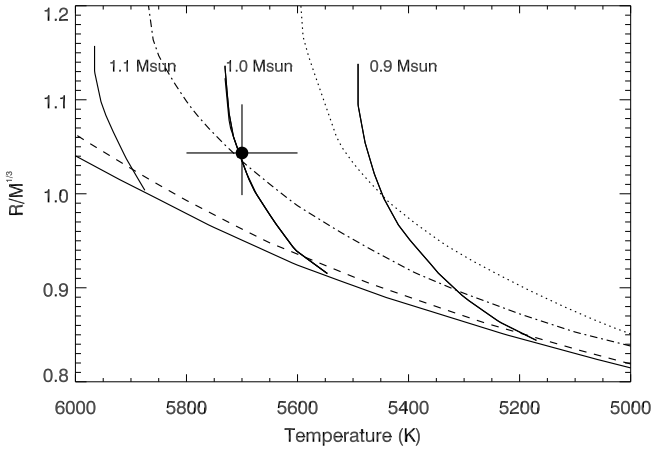}
\caption{Similar modified HR diagram than Fig. 7 but here for WASP-5. According to the models, the host star has an age of $5.4^{+4.4}_{-4.3}$ Gyr. }
\end{figure}

\begin{figure}
\label{fig:f}
\centering                     
\includegraphics[width=9cm]{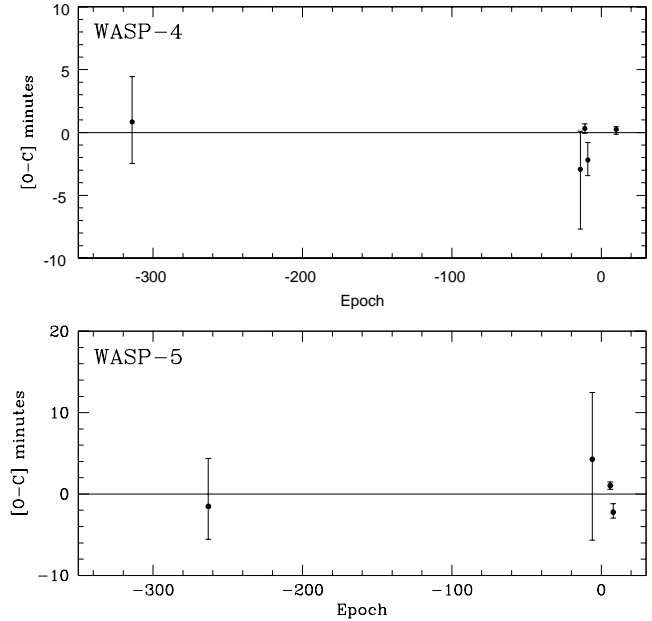}
\caption{Observed minus calculated (O-C) transit timings for the light curves included in this study. Table 2 lists the actual transit timings. The calculated timings were obtained from the ephemeris due to our combined MCMC analysis (see Table 3).}
\end{figure}

\begin{table*}
\label{tab:params}
\begin{tabular}{lcccl}
\hline
Parameter & Symbol & WASP-4 & WASP-5 & Units \\
\hline
Transit epoch (HJD) & $ T_0  $ & $ 2454383.313070^{+ 0.000045}_{- 0.000074} $ & $ 2454373.99598^{+ 0.00025} _{-0.00019} $ & days \\
Orbital period & $ P  $ & $ 1.3382324^{+ 0.0000017 }_{- 0.0000029 } $ & $ 1.6284279^{+0.0000022}_{-0.0000049} $ & days \\
Planet/star area ratio  & $ (R_p/R_s)^2 $ & $ 0.02357 ^{+ 0.00010 }_{- 0.00008 } $ & $ 0.01180^{+0.00022}_{-0.00029} $&  \\
Transit duration & $ t_T $ & $ 0.08831^{+ 0.00016}_{- 0.00021} $ & $ 0.0987^{+0.0022}_{-0.0020}$ & days \\
Impact parameter & $ b $ & $ 0.063^{+ 0.047}_{- 0.062} $ & $ 0.31^{+0.06}_{-0.28} $ & $R_*$ \\
  &    &      &  \\
Stellar reflex velocity & $ K_1 $ & $ 0.2476^{+ 0.0139}_{-0.0068} $ & $ 0.2797^{+0.0092}_{-0.0063}$ & km s$^{-1}$ \\
Centre-of-mass velocity  & $ \gamma $ & $ 57.7387 ^{+ 0.0026}_{- 0.0013} $ &  $20.0087 ^{+0.0032}_{-0.0025}$  & km s$^{-1}$ \\
Orbital semi-major axis & $ a $ & $ 0.02255 ^{+ 0.00095}_{- 0.00065} $ & $0.0267^{+0.0012} _{-0.0008}$ & AU \\
Orbital inclination & $ I $ & $ 89.35^{+ 0.64 }_{- 0.49} $ &   $86.9 ^{+2.8}_{-0.7}$ & degrees \\
Orbital eccentricity & $ e $ & 0.0 (fixed) & $0.038 ^{+ 0.026}_{-0.018}$&  \\
Longitude of periastron & $ \omega $ &  - & $0.60^{+0.47}_{-0.39}$ & rad  \\
  &    &      &  \\
Stellar mass & $ M_* $ & $ 0.85^{+ 0.11 }_{- 0.07} $ & $0.96 ^{+0.13}_{-0.09}$ & $M_\odot$ \\
Stellar radius & $ R_* $ & $ 0.873^{+ 0.036}_{- 0.027} $ & $1.029 ^{+0.056}_{-0.069}$ & $R_\odot$ \\
Stellar surface gravity & $ \log g_* $ & $ 4.487^{+ 0.019}_{- 0.015} $ &$ 4.395^{+0.043}_{-0.040}$ & [cgs] \\
Stellar density & $ \rho_* $ & $ 1.284^{+ 0.013}_{- 0.019} $ & $ 0.88 \pm 0.12 $ & $\rho_\odot$ \\
  &    &      &  \\
Planet radius & $ R_p $ & $ 1.304 ^{+ 0.054}_{- 0.042} $ & $ 1.087^{+0.068}_{-0.071}$ & $R_J$ \\
Planet mass & $ M_p $ & $ 1.21 ^{+ 0.13}_{- 0.08} $ & $1.58 ^{+0.13}_{-0.10}$ & $M_J$ \\
Planetary surface gravity & $ \log g_p $ & $ 3.212 ^{+ 0.025}_{- 0.011} $ & $3.485 ^{+0.054}_{-0.043}$ & [cgs] \\
Planet density & $ \rho_p $ & $ 0.546 ^{+ 0.039 }_{- 0.025} $ &  $  1.23 ^{+0.26}_{-0.16} $ &  $\rho_{Jup}$ \\
Planet temperature ($A=0$, F=1)  & $ T_{\mbox{eq}} $ & $ 1650 \pm 30 $ & $1706 ^{+52}_{-48}$ &  K \\
\hline\\
\end{tabular}
\caption[]{WASP-4 and WASP-5 system parameters and 1-$\sigma$ error limits derived
from MCMC analysis.}
\end{table*}

\section{Discussion}

\subsection{Ground-based photometry}

Until recently, it was considered by many that ground-based photometry could not 
reach the high cadence sub-mmag regime because of the presence 
of the atmosphere. Indeed, high frequency atmospheric noises 
(mainly scintillation) limit the precision that high SNR photometry 
can reach within small time bins.  If one is willing to compromise on the 
sampling of their photometric time-series, binning the data (or using longer
exposures) allows for getting better errors, but the obtained precision
will be finally limited by low frequency noises. To observe several times
the same planetary eclipse and to fold the photometry with the orbital
period is thus generally considered as the only option to get very
well sampled and precise eclipse light curves from the ground. 
Nevertheless, we show here that reaching the sub-mmag sub-min regime for
one eclipse is possible with a large aperture ground-based instrument. 
Photon noise and scintillation are not a concern even for small time bins when 
a rather bright transiting system like WASP-4 is monitored with a large aperture telescope like the VLT. The high standard quality of such a telescope and the excellent atmospheric conditions at Paranal  lead furthermore to a very low level of covariant noise in the differential photometry. 

Unfortunately, we report  the presence of an instrumental effect on the VLT that damaged a part of our photometry and that could be a major problem for similar programs. It was unfortunately the case for  WASP-5: the photometry is too damaged by the effect to be useful.  As mentioned in Sec, 2, the effect is probably due to the fact that  we had to turn off the active optics system of the VLT to obtain the required very large defocus. We emphasize that imaging  in this manner is not at all a standard observational mode on the VLT. We suggest for similar programs a milder level of defocus combined with the use of the active optics system. 

Recently, exquisite ground-based transit photometry for the hot Neptune GJ 436b was obtained by Alonso et al. (2008) using a different approach. They observed a transit in the H-band with the TCS telescope and its CAIN-II near-IR detector. As the red dwarf GJ 436 is very bright in the H-band (H = 6.3), the background variability is not a concern, so no dithering pattern was used and the images were severely defocused, i.e. the strategy was very similar to the one that we choose for our VLT observations. The most surprising point is that no differential photometry was used by Alonso et al. to reach such a high photometric quality. These authors explain this by the much smoother behavior of the transparency variations in the H-band compared to the visible. It is very desirable
to confirm this point by obtaining more high-quality eclipse light curves in the near-IR. Unfortunately,
this method is limited to stars that are very bright in the near-IR, and only a few are known to harbor a transiting planet (e.g. HD\,189733, HD\,209458).

\subsection{The hot Jupiters WASP-4b and WASP-5b}

The high-quality of the WASP-4 VLT $z$-band transit photometry allows a significant improvement of the precision on the impact parameter (0.063$^{+0.047}_{-0.062}$  vs 0.13$^{+0.13}_{-0.12}$ in W08) that leads to a better precision on the orbital inclination (89.35$^{+0.64}_{-0.49}$ vs 88.59$^{+1.36}_{-1.50}$ degrees in W08), but our analysis fails to give a more precise value for the planetary mass (1.21$^{+0.13}_{-0.08}$ $M_{J}$ vs 1.215$^{+0.09}_{-0.08}$ $M_J$ in W08) and even for the planetary radius (1.304$^{+0.054}_{-0.042}$ $R_J$ vs 1.416$^{+0.068}_{-0.043}$ $R_J$ in W08) despite a significantly more precise determination of the planet/star area ratio (0.02357$^{+0.00010}_{-0.00008}$ vs 0.01192$^{+0.00036}_{-0.00027}$  $R_J$ in W08). This is due to the fact that our new spectroscopy does not improve significantly our knowledge of the host star because it is now limited by the accuracy/validity of the stellar atmospheric and evolution models, and this uncertainty on the stellar parameters propagates to our final accuracy on the planet parameters. This fact is also illustrated with our results for WASP-5b: our new values for the planetary mass and radius agree well with the ones quoted in A08 (1.58$^{+0.13}_{-0.10}$ $M_{J}$ and 1.087$^{+0.068}_{-0.071}$ $R_J$ vs 1.58$^{+0.13}_{-0.08}$ $M_{J}$ and 1.090$^{+0.024}_{-0.058}$ $R_J$ in A08) but are not more accurate. This `stellar' accuracy limit relates to all transiting systems, except the putative systems (1) that could be studied by asteroseismology (see e.g. Christensen-Dalsgaard et al. 2007), allowing a more precise estimation of their age and radius, or (2) the systems that would be bright and nearby enough to allow a very precise direct determination of the stellar radius by long baseline interferometry (see e.g. Baines et al. 2008). 

With a radius measurement $R_p = 1.30^{+0.05}_{-0.04}$ $R_J$, we confirm here that WASP-4b is larger than predicted by basic models of irradiated planets (Burrows et al. 2007a, Fortney et al. 2007). For instance, the theoretical value presented by Fortney et al. (2007) for an irradiated 1.46 $M_J$ core-less planet of 4.5 Gyr orbiting a 0.02 AU from a sun-like star is only 1.17 $R_J$. WASP-4b is thus another case demonstrating that something is missing in basic models. 

Fortney et al. (2008) proposed the theoretical division of hot Jupiters into two classes based on their level of irradiation. By analogy with M-dwarfs, the `pM' class would be composed of the  planets warmer than required for condensation of Ti and V-bearing compounds, and these planets should show a stratospheric temperature inversion due to the absorption of most of the large incident flux by the high-opacity gaseous TiO and VO. The cooler planets would compose the `pL' class.  Burrows et al. (2008) proposed a similar bifurcation into two groups, but without firmly identifying TiO and VO as the high opacity gaseous compounds causing the thermal inversion. Interestingly, recent $Spitzer$ secondary eclipse measurements detected such temperature inversions for the highly irradiated planets HD~209458b and XO-1b (Burrows et al. 2007b, Knutson et al. 2008, Machalek et a. 2008). The updated estimation for the incident flux received by WASP-4b is now 1.7 10$^9$ erg s$^{-1}$ cm$^{-2}$, classing it clearly in the theoretical pM planetary class proposed by Fortney et al. Figure 10 shows the location of WASP-4b in a mass-radius diagram. The location of the other transiting planets receiving an incident flux larger than 10$^9$ erg s$^{-1}$ cm$^{-2}$ (i.e. belonging clearly to Fortney's pM class) or smaller than 5 $\times$ 10$^8$ erg s$^{-1}$ cm$^{-2}$ (i.e. belonging clearly to Fortney's pL class) is also shown for comparison. We notice that the pL planets seem to have a smaller radius than their pM counterparts having a similar mass. WASP-4b follows this tendency. This is in favor of the proposition done by Burrows et al. (2007a) that the enhanced opacity of the highly irradiated giant planets could alter their cooling history and thus be part of the solution to their `anomalously' large radius. However, WASP-5b is even more irradiated then WASP-4b (2.0 $\times$ 10$^9$ erg s$^{-1}$ cm$^{-2}$) but according to our result it is not extremely bloated ($R_p$ = 1.09 $\pm$ 0.07 $R_J$). As shown in Fig. 10, WASP-5b seems to be smaller than the other planets of similar mass that fall clearly in the  pM class. This favors the fact that the size of highly irradiated gazeous planets is not only dependent on their level of stellar irradiation but also of other factors. The planetary core mass is probably one of them. In this context, it is interesting to notice that WASP-5 is more metal-rich than WASP-4. This goes in the right direction towards the existence of a correlation between the metallicity of the parent star and the core mass of the planet  (Guillot et al. 2006, Burrows et al. 2007a, Guillot 2008).  

More transit photometry for WASP-5 is needed to confirm the possible period variability shown by our transit timing measurements. Such an apparent period variability could be due to the presence of another close-in lighter planet in the system (see e.g. Holman \& Murray 2005). A precise timing of the secondary eclipse of this planet is also desirable, because it could confirm the non-null eccentricity that we marginally deduce from our analysis. Such a non-null eccentricity could also be the sign of the presence of another planet around WASP-5. 

\begin{figure}
\label{fig:g}
\centering                     
\includegraphics[width=9cm]{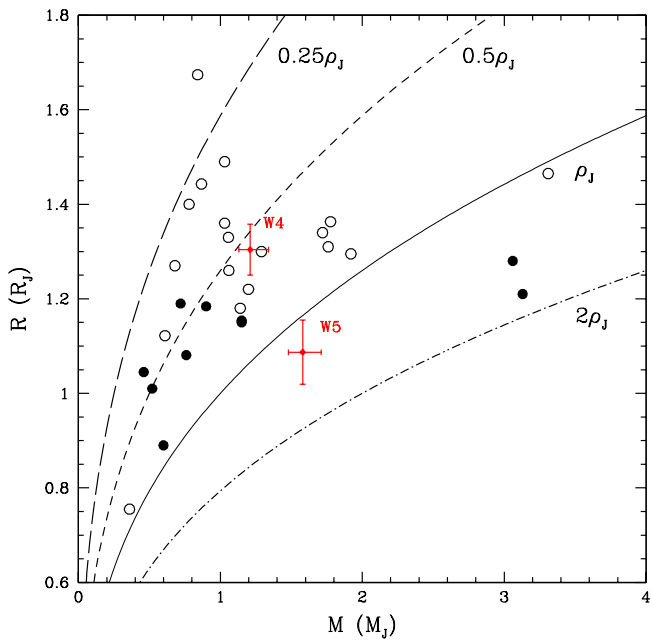}
\caption{Location of WASP-4b (W4) and WASP-5b (W5) in a mass-radius diagram. The location of the other transiting planets receiving an incident flux larger than 10$^9$ erg s$^{-1}$ cm$^{-2}$ (open circles) or smaller than 5 $\times$ 10$^8$ erg s$^{-1}$ cm$^{-2}$ (closed circles) is also shown. }
\end{figure}

\begin{acknowledgements} 
The authors thank the ESO staff on the VLT for their diligent and competent execution of the observations. E. Pompei is particularly gratefully acknowledged. We thank T. Guillot for very helpful suggestions. The anonymous referee is gratefully acknowledged for his valuable comments and suggestions.
\end{acknowledgements} 

\bibliographystyle{aa}
{}
\end{document}